\documentclass{article}

\usepackage{jheppub}
\usepackage[numbers]{natbib}

\usepackage{ulem}
\usepackage{hyperref}
\usepackage{physics}
\usepackage{todonotes}
\usepackage{enumitem}
\usepackage{amsmath,amssymb,gensymb}
\usepackage{cancel}

\title{
Axion production via trapped misalignment from Peccei-Quinn symmetry breaking
}
\author{Luca Di Luzio$^{a}$,}
\author{Philip Sørensen$^{a,b}$}
\affiliation[a]{Istituto Nazionale di Fisica Nucleare (INFN), Sezione di Padova, \\ Via F. Marzolo 8, 35131 Padova, Italy}
\affiliation[b]{Dipartimento di Fisica e Astronomia `G.~Galilei', Universit\`a di Padova, \\ Via F. Marzolo 8, 35131 Padova, Italy}
\emailAdd{luca.diluzio@pd.infn.it}
\emailAdd{philip.soerensen@pd.infn.it}

\abstract{The Peccei-Quinn (PQ) symmetry does not need to be exact, 
and even a tiny source of PQ breaking 
not aligned with the QCD anomaly 
might have 
significant phenomenological implications.  
In this study, we examine the effects of a general class of PQ-breaking operators 
on the axion cosmological production via misalignment, 
focussing on both temperature-dependent and independent PQ-breaking potentials. 
In particular, we show that a variant of the 
trapped misalignment mechanism can 
delay the onset of axion oscillation, 
leading to an axion dark matter 
window with $m_a \gg 10^{-5}$ eV. 
This scenario is testable through various experimental approaches, including standard axion haloscopes and helioscopes, as well as searches for electric dipole moments and axion-mediated forces.}

\newcommand{\mplanck}{m_{\rm Pl}}

\newcommand{\TQCD}{T_{c}}
\newcommand{\Ttoday}{T_0}

\newcommand{\deltaPQV}{\delta_{\rm \cancel{\rm PQ}}}

\newcommand{\thetaVEV}{\theta_{\rm eff}}
\newcommand{\aini}{a_{\rm ini}}
\newcommand{\thetaini}{\theta_{\rm ini}}
\newcommand{\Tosc}{T_{\rm osc}}

\newcommand{\VThermalGG}{V_{\rm th,\,GG}}

\newcommand{\VThermalfermion}{V_{\rm th,\,f}}
\newcommand{\VThermalgeneral}{V_{\rm th}}

\newcommand{\maThermalgeneral}{m_{a,\rm th}}
\newcommand{\maThermalFermions}{m_{a,\rm th,f}}

\newcommand{\maPQV}{m_{a,\rm \,\cancel{\rm PQ}}}

\newcommand{\thetaTrap}{\theta_{\rm release}}
\newcommand{\TTrap}{T_{\rm release}}
\newcommand{\chiQCD}{\chi_{\rm QCD}}

\newcommand{\rhoDMToday}{\rho_{\rm DM,\, today}}
\newcommand{\rhoaToday}{\rho_{a,\,\rm today}}
\newcommand{\LambdaPQV}{\Lambda_{\rm \cancel{\rm PQ}}}
\newcommand{\lambdaPQV}{\lambda_{\rm \cancel{\rm PQ}}}
\newcommand{\LambdaPrime}{\Lambda'}
\newcommand{\LambdaUV}{\Lambda_{\rm UV}}
\newcommand{\VPQV}{V_{\rm \cancel{\rm PQ}}}
\newcommand{\LPQV}{\mathcal{L}_{\rm \cancel{\rm PQ}}}
\newcommand{\VQCD}{V_{\rm QCD}} 
\newcommand{\dmfraction}{r_{\rm DM}}

\newcommand{\ncol}{n_{\rm col}}
\newcommand{\alphaStrong}{\alpha_{\rm s}}

\newcommand{\U}{{\rm U}}

\renewcommand{\(}{\left(}
\renewcommand{\)}{\right)}

\newcommand{\beq}{\begin{equation}}
\newcommand{\eeq}{\end{equation}}

\begin{document}
\maketitle

\newpage

\section{Introduction}
\label{sec:intro}

The Peccei-Quinn (PQ) solution to the strong CP problem 
\cite{Peccei:1977hh,Peccei:1977ur}
requires a global $\U(1)_{\rm PQ}$ symmetry, that is 
explicitly broken by the 
Quantum Chromodynamics (QCD) anomaly. 
The axion field, which arises as the Goldstone mode 
of the spontaneously broken $\U(1)_{\rm PQ}$ \cite{Weinberg:1977ma,Wilczek:1977pj}, 
also 
provides 
an excellent dark matter (DM) candidate,
based e.g.~on the so-called axion misalignment 
production mechanism \cite{Dine:1982ah,Abbott:1982af,
Preskill:1982cy}. 
The 
$\U(1)_{\rm PQ}$,
being a global symmetry, 
does not need to be exact.
In fact, it is explicitly broken by the QCD anomaly,
and it is expected to be broken further
by ultraviolet (UV) physics, 
at the very least by quantum gravity effects 
\cite{Barr:1992qq,Holman:1992us,Kamionkowski:1992mf}.
While the explicit breaking from QCD is guaranteed to generate a minimum which conserves CP~\cite{Vafa:1983tf}, 
other sources of PQ breaking would in general not be aligned to this minimum.
This implies that the axion 
vacuum expectation value (VEV)
could be displaced from the minimum at zero. 
However,
a large axion VEV would reintroduce the CP-violating effects that the axion was introduced to resolve and it is strongly constrained by the
non-observation of the  
electric dipole moment (EDM) of the neutron (nEDM) \cite{Abel:2020pzs}. Therefore, 
PQ-breaking  
potentials must 
be strongly subdominant 
today with respect
to the axion potential from QCD.

However, a source of PQ breaking might still leave an
imprint in axion cosmology. As it was studied in Ref.~\cite{Jeong:2022kdr}, 
PQ-breaking effects,
assumed to be temperature-independent,
become relevant at high temperatures, 
$T \gg T_c \approx 150$ MeV, for which the standard QCD axion potential is suppressed. 
Hence, new PQ-breaking sources 
beyond QCD
may affect the axion field evolution 
in the early universe and modify 
(by either delaying or expediting)
the time 
when the axion starts to oscillate, 
thus 
affecting 
(by either enhancing or suppressing)
the axion relic 
abundance from 
misalignment.
In Ref.~\cite{Jeong:2022kdr} it was found that the QCD axion could account for all the observed DM with masses well above the range expected from the standard misalignment mechanism, i.e.~$m_a \gg 10^{-5}$ eV. This was achieved with PQ-breaking effects which delay damping by realizing the so-called trapped misalignment \cite{Higaki:2016yqk,Kawasaki:2017xwt,DiLuzio:2021pxd,DiLuzio:2021gos,Nakagawa:2020zjr}.
However, in the framework presented there, a tuning of at least $10^{-3}$ in the alignment of the PQ-breaking potential was necessary to remain compatible with nEDM constraints. Fundamentally, this need arises because the dimensionless axion VEV today can be at most $10^{-10}$ and the QCD axion potential at the epoch relevant for
misalignment is suppressed only by a factor of about $10^{-7}$, leaving the residual factor of $10^{-3}$ unaccounted for.

In this work, we 
extend the analysis of Ref.~\cite{Jeong:2022kdr} 
by taking into account a new class of PQ-breaking operators, 
recently considered in Refs.~\cite{Zhang:2022ykd,Zhang:2023gfu}, 
in which the 
axion
couples directly 
to Standard Model (SM) fields. 
Since the latter belong to the 
early universe thermal bath, 
a remarkable consequence of those 
PQ-breaking operators 
is the generation of a temperature-dependent 
axion potential which {\it grows} with $T$, 
thus becoming
even more relevant at high temperatures, 
when compared to the suppressed QCD axion potential. 
The additional temperature dependence of the 
PQ-breaking potential can then overcome the residual hierarchy faced in Ref.~\cite{Jeong:2022kdr} and realize axion DM for $m_a \gg 10^{-5}$ eV without tuning.

Our paper is structured as follows: In section \ref{sec:relevance}, we motivate the relevance of PQ-breaking effects for misalignment and highlight the relevant hierarchies and temperature scales. In section \ref{sec:T-independent}, we briefly review the results of \cite{Jeong:2022kdr} and study trapped misalignment from a temperature-independent 
PQ-breaking potential. In section \ref{sec:Tdeppotentials}, we go beyond \cite{Jeong:2022kdr} by 
analyzing the effects of the PQ-breaking potentials 
introduced
in \cite{Zhang:2022ykd,Zhang:2023gfu} and 
showing how trapped misalignment can be realized without tuning. In this section, we also discuss relevant constraints and 
how 
such a  
relatively heavy axion DM 
scenario, with $m_a \lesssim 0.1$ eV, 
might be testable in future 
experiments, 
including standard axion haloscopes and helioscopes, 
as wells as 
searches for 
EDMs and axion-mediated forces.

\section{Relevance of PQ breaking for misalignment} 
\label{sec:relevance}

In the following, we will refer 
to the 
standard QCD axion and discuss how 
general sources of 
PQ breaking 
might allow for interesting effects at the temperatures relevant for misalignment.
First, consider the temperature dependence of the QCD axion potential, which can be written as 
\begin{gather}
	\VQCD(T) = - \chiQCD(T) \cos(\theta) \, , 
\end{gather}
where $\chiQCD(T) = f_a^2 m_a^2(T)$  is the topological susceptibility at finite temperature.  
The temperature dependence of the axion mass can be extracted from lattice QCD simulations (see e.g.~\cite{Borsanyi:2016ksw}), and for our purposes it can be approximated 
as
\begin{gather}
 m_a^2(T) = m_a^2 \, \min\left[ \left(\frac{T}{\TQCD}\right)^{-2b} ,\ 1\right]\label{eq:ma(T) lattice fit} ~ , 
\end{gather}
where $\TQCD = 150 $ MeV and $b=3.92$. In our notation, $m_a$ and $\chiQCD$ refer to their zero-temperature values when written without explicit $T$-dependence.
In the early universe, $m_a(T)$ competes against an effective friction term arising from the cosmological expansion, and the equation of motion (EOM) 
of the axion field 
in the harmonic approximation
reads
\begin{gather}
	\ddot{a}+3 H\dot{a}+m_a^2(T) a = 0 \, ,\label{eq:axion EOM}
\end{gather}
where $H$ is the Hubble parameter.  At early times, when the friction term (referred to as Hubble friction) dominates, the system is over-damped and the axion is effectively frozen at some initial value $\aini = \theta_{\rm ini} f_a$, where $\thetaini \in [-\pi,\pi)$ is the initial misalignment angle. 
If the spontaneous breaking of the PQ symmetry takes place after inflation, then each patch of the universe may select a different value and 
we expect $\aini$ to take an averaged value. 
In such post-inflationary scenarios, topological defects 
such as axion strings and domain walls
also play a major role. However, if PQ breaking takes place before the end of inflation, then inflation generically selects a single randomly chosen value 
for the initial condition, such that we naturally expect $\thetaini \sim \mathcal{O}(1)$. In this work, we focus on such pre-inflationary scenarios in which topological defects are absent.

Eventually, at some temperature $\Tosc$ the axion starts to oscillate once $\VQCD$ overcomes Hubble friction, i.e.~when
\begin{gather}
	m_a(\Tosc) \approx 3 H(\Tosc) \, . \label{eq:hubble friction release}
\end{gather}
After the mass term dominates, the axion EOM becomes a damped harmonic oscillator which behaves as DM. In this limit, the EOM eq.~\eqref{eq:axion EOM} can be solved by a WKB approximation which yields $a \propto m_a^{-1/2}(T)\ R^{-3/2} $, where $R$ is the scale factor. Assuming that this WKB evolution takes effect immediately after $\Tosc$, oscillations starting with an amplitude of $\aini$ lead to a present-day DM relic of the form
\begin{gather}
	\rhoaToday \approx \frac{1}{2}\chiQCD\thetaini^2 \frac{m_a(\Tosc)}{m_a}\frac{g_{*s}(\Ttoday)}{g_{*s}(\Tosc)}\left(\frac{\Ttoday}{\Tosc}\right)^3~,\label{eq:present day relic of Tosc}
\end{gather}
where $g_{*s}(T)$ is the number of relativistic degrees of freedom in entropy and $\Ttoday$ is the present-day temperature. Anharmonic effects may contribute an $\mathcal{O}(1)$ factor which we neglect here for simplicity.
Because $ \chiQCD = f_a^2 m_a^2 \approx \left(76\text{ MeV}\right)^4 $ is fixed for any QCD axion, the present-day observed DM density is found for a specific amount of redshift corresponding to a unique oscillation temperature. For the axion  to make up a fraction $\dmfraction$ of the observed DM abundance, the axion must start to oscillate at
\begin{gather}
	\Tosc \approx 950 \text{ MeV } \dmfraction^\frac{-1}{3+b} \thetaini^\frac{2}{3+b}.\label{eq:ToscDM}
\end{gather}
In the conventional scenario, in which the condition eq.~\eqref{eq:hubble friction release} controls $\Tosc$, $\thetaini$ must be tuned to compensate for the failure to hit $\Tosc \sim 950$ MeV if the axion mass deviates significantly from $m_a \sim 10^{-5}$ eV. However, if we introduce additional dynamics which can modify the starting temperature of axion oscillations, then such an additional degree of freedom can be used to fix the condition eq.~\eqref{eq:ToscDM} even for axion masses far from $m_a \sim 10^{-5}$ eV. This is the underlying principle of approaches such as \textit{trapped misalignment} \cite{Higaki:2016yqk,Kawasaki:2017xwt,DiLuzio:2021gos,Nakagawa:2020zjr} and \textit{kinetic misalignment} \cite{Co:2019jts,Chang:2019tvx}, which have been recently discussed in the literature.

The central question of this work is how PQ-breaking potentials
can implement trapped misalignment and thus be used to motivate axion DM across a non-standard parameter space. Naively, one might expect this not to be possible given that any source of PQ breaking must be small enough not to generate an axion VEV, 
$\thetaVEV \equiv \langle a \rangle / f_a$,
that overshoots the nEDM limit
of $|\thetaVEV| \lesssim 10^{-10}$. 
This means that any PQ-breaking potential must be far subdominant to $\VQCD$ today. However, because of the temperature dependence of the QCD axion potential, eq.~\eqref{eq:ma(T) lattice fit}, this need not always have been the case.

In particular, the hierarchy between $\TQCD \sim 150 $ MeV and $\Tosc\sim 950$ MeV generates a suppression of the QCD potential of 
\begin{gather}
    \frac{\VQCD(\Tosc)}{\VQCD(T=0)} \approx \left( \frac{\TQCD}{\Tosc} \right)^{2b} \sim 5 \times 10^{-7},
\end{gather}
which makes it much easier for PQ-breaking potentials to compete near $\Tosc$. Nevertheless, as we will show explicitly in section \ref{sec:nEDMtuning},
this implies that a temperature-independent PQ-breaking potential 
in the absence of tuning
would not be able to compete at $\Tosc\sim 950$ MeV, as this would require a suppression of $10^{-10}$ in order to overcome nEDM limits.

Which temperature then is high enough for PQ breaking to dominate? 
If we give up on the assumption that the axion makes up all of the observed DM, then we can consider misalignment at higher temperatures where the QCD potential is further suppressed.
Because the misalignment temperature $\Tosc$ is controlled by $m_a$ through the condition $m_a(\Tosc) \approx 3H(\Tosc)$, pushing $\Tosc$ to higher temperatures requires us to consider larger axion masses.
In order for the QCD potential to be suppressed by a factor of $10^{-10}$, such that constant-temperature PQ-breaking can dominate, requires axion masses of 
\begin{gather}
	m_a \gtrsim 5 \times 10^{-3} \text{ eV} ~. \label{eq:ma lower bound for CPV effect}
\end{gather}
Of course, if misalignment takes place at such temperatures, then DM would be under-produced according to eq.~\eqref{eq:ToscDM}. Nevertheless, the fact that PQ-breaking effects can play a role means that interesting phenomenology might arise. We review this constant-temperature scenario in the next section. Furthermore, it is clear that the competitiveness of the PQ-breaking potential can be drastically improved if this grows with $T$, 
a case that will be discussed 
in section \ref{sec:Tdeppotentials}.


\clearpage
\section{Axion DM production from \texorpdfstring{$T$}{T}-independent PQ-breaking potentials}\label{sec:T-independent}

We now discuss the scenario of a temperature-independent PQ-breaking potential in more detail. 
To keep the discussion general, we will parameterize the potential as
\begin{gather}
		\VPQV = -\LambdaPQV^4\cos(n\theta+\deltaPQV) ~ ,\label{eq:constant T potential, general form}
\end{gather}
where $\theta = a /f_a$ is the adimensional axion field,
$\LambdaPQV$ describes the amplitude of the 
PQ-breaking potential, $n$ is an integer and 
$\deltaPQV$ is a generic phase that parametrizes the misalignment of the minimum from that of the QCD axion potential.

\begin{figure}
    \centering
    \includegraphics[width=0.8\textwidth]{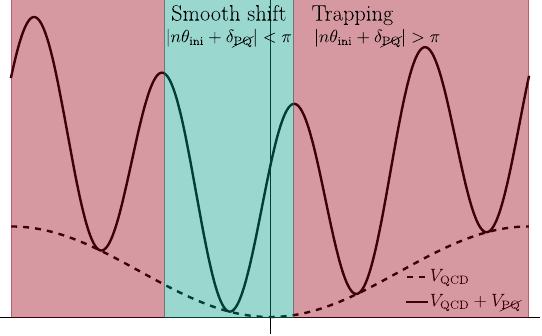}
    \caption{Illustration of the two principal regimes defined by eqs.~\eqref{eq:smooth shift regime} and \eqref{eq:trapping regime}. If $\thetaini$ lies in the minimum of $\VPQV$ closest to the origin (defined by the minimum of $\VQCD$), then $\theta$ will adiabatically shift towards zero. This defines the smooth shift regime. On the contrary, if $\thetaini$ lies in any other minimum, then $\theta$ is separated from the origin by a potential barrier resulting in a trapped scenario.}
    \label{fig:twoRegions}
\end{figure}

A potential of the form in eq.~\eqref{eq:constant T potential, general form} can be generated from 
UV sources of PQ-breaking, 
possibly related to quantum gravity effects 
(see e.g.~\cite{Kallosh:1995hi}).
Such effects can be encoded 
through
an effective operator of the type 
\begin{gather}
	V\supset \label{eq:VPQV}
	- e^{i \deltaPQV} \frac{\phi^n}{\LambdaUV^{n-4}}+\text{h.c.} 
	\, ,  
\end{gather}
with $n>4$. 
Here,    
$\Lambda_{\rm UV}$ is a UV scale
and $\phi=\frac{f_a}{\sqrt{2}} e^{i\theta}$ is a complex scalar,  
with the radial mode integrated out and 
the angular mode
given by the 
adimensional axion field. 
When written in terms of $\theta$, an operator of the form in eq.~\eqref{eq:constant T potential, general form} arises with an amplitude of
\begin{equation}
    \LambdaPQV^4 = 
    \frac{1}{2^{n/2-1}} \frac{f_a^n}{\LambdaUV^{n-4}} \, .
\end{equation}
While 
this provides a canonical example,
in this work we remain agnostic about the source of PQ breaking and work with the generic potential in eq.~\eqref{eq:constant T potential, general form}.
The impact of such a $T$-independent PQ-breaking potential was investigated in Ref.~\cite{Jeong:2022kdr}.
If the PQ-breaking potential dominates 
with respect to
the QCD axion potential prior to misalignment, which in the absence of tuning in the  CP-violating phase
$\deltaPQV$ is subject to the condition eq.~\eqref{eq:ma lower bound for CPV effect}, then the PQ-breaking potential modifies the resulting relic density.
As pointed out by the authors of Ref.~\cite{Jeong:2022kdr}, the impact of such a PQ-breaking potential can be classified into two distinct regimes:
\begin{align}
	\text{Smooth shift regime:} \qquad \abs{n\thetaini + \deltaPQV} &< \pi  ~,  \label{eq:smooth shift regime}\\
	\text{Trapping regime:} \qquad \abs{n\thetaini + \deltaPQV} &> \pi  ~. \label{eq:trapping regime} 
\end{align}
These conditions are visualized in figure \ref{fig:twoRegions}.
In the first regime, smooth shift, the main impact of the PQ-breaking potential is to trigger oscillations earlier than what would have been achieved by $\VQCD$ alone. This results in increased redshift and therefore a lower present-day DM abundance. 
Such a scenario takes place when $\thetaini$ selects the minimum of $\VPQV$ closest to the minimum of $\VQCD$, which permits an adiabatic transition between the two minima. However, if $\thetaini$ selects a minimum of $\VPQV$ further from the minimum of $\VQCD$, then a potential barrier separates the two minima. This potential barrier prevents an adiabatic transition and results in trapped misalignment. In this scenario, the field first settles into and tracks the initial, false minima of $\VPQV$. The oscillations in this initial misalignment process are damped out by Hubble friction, and we can consider the field to be at rest at the time-dependent minimum. Then, at some transition temperature, $\TTrap$, the potential barrier eventually disappears and the field suddenly starts oscillating around the late-time minimum of $\VQCD$, $\theta=0$. This essentially restarts oscillations at a temperature below the temperature at which oscillations would have taken place given only $\VQCD$, resulting in less redshift and a larger present-day DM abundance.

As discussed in the previous section, general arguments lead us to conclude that temperature-independent PQ-breaking effects can only become relevant 
in the regime in which the misalignment mechanism 
underproduces axion DM, see eq.~\eqref{eq:ma lower bound for CPV effect}. Therefore, we expect more interesting phenomenology from the trapped regime, which we focus on in the remainder of this work.

\subsection{Trapping and release}
If the axion starts oscillating in a local minimum that is not the closest one to the CP-conserving origin, then there will be a potential barrier which prevents the transition from taking place in an adiabatic way. Instead, the axion will be trapped in the false minimum until the potential barrier disappears. 
The disappearance of the potential barrier requires that
\begin{gather}
	\pdv{V}{\theta}=0 \quad \qq{and} \quad \pdv[2]{V}{\theta}=0 \, ,
\end{gather}
where $V=\VQCD+\VPQV$. With the PQ-breaking potential given by 
eq.~\eqref{eq:constant T potential, general form}, these two conditions can be rewritten into the form
\begin{gather}
	\tan{\thetaTrap} = \frac{1}{n} \tan(n\thetaTrap+\deltaPQV)\, ,\label{eq:thetaTrap equation} \\
	\frac{\TTrap}{\TQCD} = \left(\frac{-\chiQCD \sin \thetaTrap }{n \LambdaPQV^4 \sin(n \thetaTrap +\deltaPQV)} \right)^{\frac{1}{2b}}\label{eq:TTrap full equation} \, .
\end{gather}
Taken in isolation, eq.~\eqref{eq:thetaTrap equation} has $2(n-1)$ solutions. $n-1$ of these solutions also solve \eqref{eq:TTrap full equation}, corresponding to transitions from the $n-1$ minima of $\VPQV$ with trapping solutions.
As it was pointed out in \cite{Jeong:2022kdr}, 
$\thetaTrap$ does not depend on the amplitude of the PQ-breaking potential.

Eq.~\eqref{eq:TTrap full equation}, given a suitable solution of eq.~\eqref{eq:thetaTrap equation}, governs the temperature at which the transition takes place. These equations do not have exact analytical solutions, but by discarding the angular dependence in eq.~\eqref{eq:TTrap full equation} we can approximate $\TTrap$ as 
\begin{gather}
	\TTrap \approx \TQCD \left(\frac{\chiQCD}{n\LambdaPQV^4}\right)^{\frac{1}{2b}} = 1.3\text{ GeV } \left(\frac{n^{1/4}\LambdaPQV}{10^{-3}\text{ GeV}}\right)^{-0.51}.\label{eq:TTrap simplified}
\end{gather} 
In the absence of tuning, the error arising from discarding the angular dependence corresponds to an $\mathcal{O}(1)$ factor in $\TTrap$. Interestingly, this simplified solution can be also 
obtained by comparing $m_a(T)$ with an effective mass defined as
\begin{gather}
	\maPQV^2\equiv \frac{1}{a}\pdv{\VPQV}{a}\, ,\label{eq:maEff constant T}
\end{gather}
with $m_a(T)$ and $\maPQV$ being evaluated in the minima of their respective potentials. These effective masses are defined such that the EOM, $\ddot{a}+3H\dot{a}+V'=0$, matches the standard form $\ddot{a}+3H\dot{a}+ \maPQV^2 \, a = 0$.
The simplified solution eq.~\eqref{eq:TTrap simplified} is then approximated by comparing the masses as $m_a(\TTrap)\approx\maPQV$ and then solving for $\TTrap$, which 
allows us to better understand the relative importance of the two potentials.

\subsection{Axion DM relic abundance}\label{sec:dark matter for constant T}
After the axion is released from the false minimum at $\TTrap$, the field oscillates in $\VQCD$ as it would in standard misalignment. As before, we can then assume an instantaneous entry to the WKB regime to obtain the present-day axion relic abundance from eq.~\eqref{eq:present day relic of Tosc}, with $\TTrap$ replacing $\Tosc$. To simplify the expression, we approximate the $g_{*s}(\TTrap)\approx g_{*s}(1\, \text{GeV})\approx 69$, which yields
\begin{gather}
	\rhoaToday \approx 2.9\times 10^{-2} \chiQCD \left(\frac{\TQCD}{\TTrap}\right)^{b} \left(\frac{\Ttoday}{\TTrap}\right)^3 \thetaTrap^2 \, .\label{eq:rhoaToday of Ttrap}
\end{gather}
Applying the approximate solution for $\TTrap$, the present-day relic takes the form
\begin{gather}
	\rhoaToday \approx 2.9\times 10^{-2} (\chiQCD)^{1-\frac{3}{b}} \left(\frac{\Ttoday}{\TQCD}\right)^3 n^{\frac{3+b}{2b}} \thetaTrap^2 \LambdaPQV^{2+\frac{6}{b}} \, . 
 \label{eq:rhoaToday constant T}
\end{gather}
As it is clear from the above discussion, the effect of the PQ-breaking potential is to delay oscillations. This reduces the amount of redshift the DM relic experiences between $\TTrap$ and today and thus increases the predicted DM relic density. The effect grows with increasing $\LambdaPQV$.

\subsection{nEDM limits and tuned PQ-breaking scenario}
\label{sec:nEDMtuning}

The size of $\LambdaPQV$ is ultimately limited by nEDM constraints.
Unless the misalignment angle $\deltaPQV$ is tuned to be small, this requires $\LambdaPQV^4 \ll \chiQCD$. Specifically, $\VPQV$ generates a VEV of size
\begin{gather}
\thetaVEV \approx - \frac{n \LambdaPQV^4}{\chiQCD} \sin \deltaPQV \, ,
\end{gather}
where we expanded in the limit $\thetaVEV \ll 1$ and $\LambdaPQV^4 \ll \chiQCD$. nEDM constraints restrict this VEV to $|\thetaVEV| \lesssim 10^{-10}$, which implies an upper bound of 
\begin{gather}
\LambdaPQV \lesssim 2.4 \times 10^{-4} \text{ GeV } \left(n \sin\deltaPQV \right)^{-1/4} \, . \label{eq:LambdaPQV limit from nEDM}
\end{gather}
Plugging this into eq.~\eqref{eq:rhoaToday constant T} reveals that this limit restricts the axion DM relic to at most
\begin{gather}
\label{eq:analapprox}
	\frac{\rhoaToday}{\rhoDMToday} \lesssim 5\times 10^{-4} \frac{\thetaini^2}{\sin^{0.88}(\deltaPQV)} \, , 
\end{gather}
which rules out the possibility of trapped misalignment with a temperature-independent potential to generate the observed DM abundance, unless 
the PQ-breaking phase is tuned down to $\deltaPQV \lesssim 10^{-3}$. 

\subsection{Numerical analysis}
\begin{figure}[t!]
	\centering
	\includegraphics[width=0.8\textwidth]{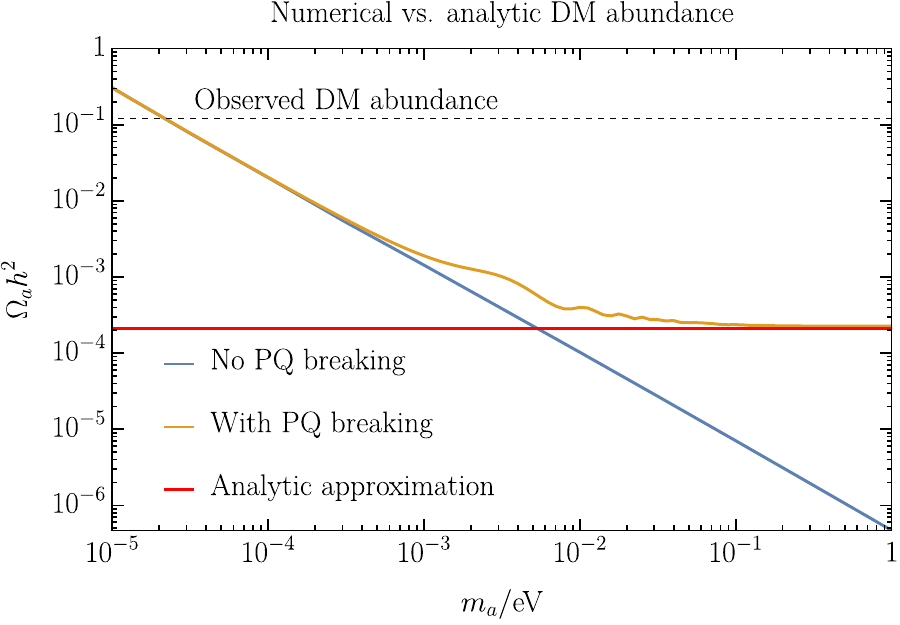}
	\caption{Numerical solution of the axion abundance, showing the smooth transition from the regime of standard misalignment into the trapped regime in which the axion relic becomes independent of $m_a$ 
    The solution assumes that $\LambdaPQV$ saturates the nEDM bound (eq.~\eqref{eq:LambdaPQV limit from nEDM}) and an $\mathcal{O}(1)$ factor has been included to match the relic predicted by eq.~\eqref{eq:rhoaToday constant T} to the asymptotic behaviour of the numerical result. The model parameters used in this example are $n=3$, $\thetaini=2$ and $\deltaPQV =2.5$.}
    \label{fig:axionrelicnum} 
\end{figure}

We have implemented a numerical solution to confirm the 
analytic approximation in eq.~\eqref{eq:analapprox}. 
The numerical analysis,  
exemplified in figure \ref{fig:axionrelicnum}, 
shows that once $\LambdaPQV$ is assumed to saturate the nEDM limit then the axion relic density begins to significantly deviate from the standard result for $m_a \gg 10^{-3}$ eV, which confirms the expectation from eq.~\eqref{eq:ma lower bound for CPV effect}. Furthermore, for $m_a \gg 10^{-3}$ eV the relic abundance enters a regime in which the relic abundance becomes independent of $m_a$. The relic abundance in this regime agrees with our analytic estimate for $\rhoaToday$ up to an $\mathcal{O}(1)$ factor, which can be attributed to anharmonic effects and the failure of oscillations to instantaneously enter the WKB regime after the release at $\TTrap$. 

\clearpage
\section{Axion DM production from \texorpdfstring{$T$}{T}-dependent 
PQ-breaking
potentials}
\label{sec:Tdeppotentials}

In the previous section, we have seen how the rapid decay of $\VQCD$ for $T\gtrsim \TQCD$ allows for a $T$-independent PQ-breaking potential to 
impact the misalignment mechanism. However, although a trapping mechanism allows $\VPQV$ to dominate and postpone the onset of oscillations, the effect is insufficient to account for axion DM in the trapping regime. 
The main cause for this limitation is that the suppression of $\VQCD$ is insufficient,
barring a tuning of $\deltaPQV$,
to overturn the nEDM-enforced hierarchy between $\VQCD$ and $\VPQV$ if the latter potential does not depend on $T$. To escape this limitation, we will now investigate a class of PQ-breaking operators 
recently considered in \cite{Zhang:2022ykd,Zhang:2023gfu}:
\begin{gather}
	\label{eq:phiOSM}
	\left(\frac{\phi}{\LambdaUV}\right)^n \mathcal{O}_{\rm SM} \, ,
\end{gather}
where $\mathcal{O}_{\rm SM}$ is a renormalizable SM
operator. The advantage of such a term is that it generates a temperature-dependent thermal potential, so that
$\VPQV$ grows with temperature. 
In \cite{Zhang:2022ykd,Zhang:2023gfu} the operator \eqref{eq:phiOSM} was considered with only $n=1$. We have here generalized eq.~\eqref{eq:phiOSM} to $n\geq 1$, since $n=1$ only offers the smooth shift solution which would result in a strongly suppressed DM relic.

In this section, we investigate the possibilities 
offered by the above scenario. First, we proceed to generalize the results of section \ref{sec:T-independent} to a general, temperature-dependent potential. Then, we present specific implementations of 
eq.~\eqref{eq:phiOSM}. Finally, we discuss the axion VEV,
the CP-violating axion interactions with SM matter fields
and the associated constraints,  
before we summarize our results. 

\subsection{Trapped misalignment 
in the presence of \texorpdfstring{$T$}{T}-dependent PQ-breaking potentials}

Before coming to specific examples of axion potentials 
growing with $T$,  
let us discuss some general features of 
trapped misalignment for 
such thermal potentials.
We will parameterize these as 
\begin{gather}
	\VThermalgeneral = -\LambdaPQV^4(T) \cos(n\theta + \deltaPQV) \, , 
 \quad \qq{where} \quad \LambdaPQV^4(T) = \lambdaPQV^4 \left(\frac{f_a}{\sqrt{2}\LambdaUV}\right)^n T^q \, ,  \label{eq:Vthermal general}
\end{gather}
and $q$ is a positive power. 
At high temperatures, such a potential can dominate with respect to the QCD potential. With the replacement $\LambdaPQV \to \LambdaPQV(T)$ many of the results from the previous section can be reapplied. In particular, for $n>1$ it is possible to have trapped solutions with the release conditions 
set by eqs.~\eqref{eq:thetaTrap equation}-\eqref{eq:TTrap full equation}.
It is also possible to generalize the approximate solution in eq.~\eqref{eq:TTrap simplified}, which becomes 
\begin{gather}
	\TTrap \approx \TQCD \left(\frac{\chiQCD}{n\LambdaPQV^4(\TTrap)}\right)^{\frac{1}{2b}}. \label{eq:TTrap simplified of T}
\end{gather}
By solving \eqref{eq:TTrap simplified of T} for $\TTrap$, we can then find the release temperature for a thermal potential with arbitrary $T$ dependence. If we factor out the $T$ dependence as in 
eq.~\eqref{eq:Vthermal general}, then the general solution takes the form
\begin{gather}
	\TTrap \approx \left(2^{\frac{n}{4 b}} n^{-\frac{1}{2 b}} \lambdaPQV^{-2/b} m_a^{1/b} f_a^{-\frac{n-2}{2 b}} \TQCD  \LambdaUV^{\frac{n}{2 b}}\right)^{\frac{2 b}{2 b+q}}. \label{eq:Ttrap simple solution for any T}
\end{gather}
The present-day axion relic is then found by plugging eq.~\eqref{eq:Ttrap simple solution for any T} into eq.~\eqref{eq:rhoaToday of Ttrap}. To produce 
an axion abundance $\rhoaToday = \dmfraction \, \rhoDMToday$ requires a specific value of $\LambdaUV$, which is
\begin{gather}
\LambdaUV \approx \frac{0.60^{\frac{2 b+q}{n}}}{\sqrt{2}} f_a n^{1/n} \lambdaPQV^{4/n} \TQCD^{\frac{b (q-6)}{(b+3) n}} \Ttoday^{\frac{3 (2 b+q)}{(b+3) n}} \chiQCD^{\frac{b+q-3}{(b+3) n}} (\dmfraction \, \rhoDMToday)^{-\frac{2 b+q}{b n+3 n}} \abs{\thetaini}^{\frac{2 (2 b+q)}{(b+3) n}}. \label{eq:LambdaUV DM sol, general}
\end{gather}
Hence, given an operator which gives rise to a potential of the form of eq.~\eqref{eq:Vthermal general}, we can find a solution for the value of the scale $\LambdaUV$ which reproduces DM simply by reading the parameter $\lambdaPQV$ from the definition in eq.~\eqref{eq:Vthermal general}.

\begin{figure}
	\centering
	\includegraphics[width=0.85\textwidth]{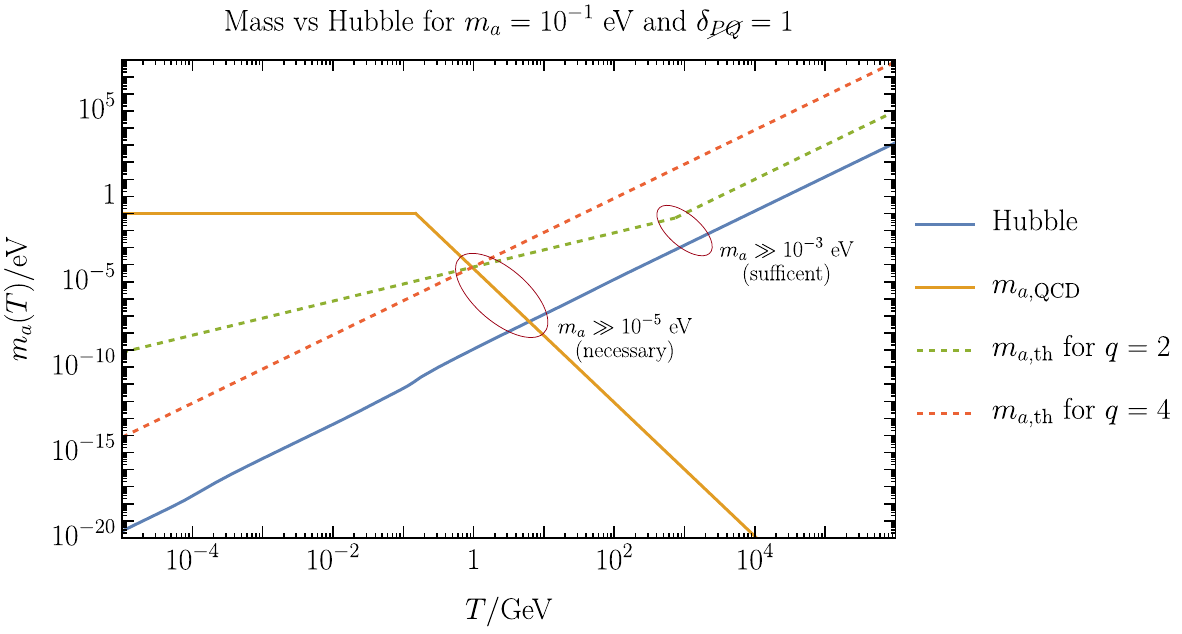}
	\caption{The effective masses generated by thermal potentials $\propto T^2$ (green-dashed) or $\propto T^4$ (red-dashed) as compared to the QCD-generated axion mass $m_a(T)$ and the Hubble parameter, $H(T)$. The thermal potential dominates for $T\gg$ GeV and the transition between $m_a(T)$ and the thermal masses takes place at $T\sim 950$ MeV, effectively postponing oscillations in the late-time minima. The fermion Yukawa transitions to a $T^4$ evolution above $T\sim v$, which accounts for the kink in the green dashed line. The features ensured by eqs. \eqref{eq:maMin, strict} and \eqref{eq:maMin, loose} are highlighted by red circles.}
	\label{fig:evolutionPlotCombined}
\end{figure}

Interestingly, regardless of the origin of PQ breaking,
once the DM abundance is fixed according to eq.~\eqref{eq:LambdaUV DM sol, general}, the amplitude of the 
PQ-breaking potential is fixed 
by the condition $\VPQV(\TTrap)=\VQCD(\TTrap)$.
Because the overall amplitude at $\TTrap$ is fixed by the DM requirement, the evolution of the PQ-breaking potential depends just on the scaling power $q$ (and implicitly on $m_a$ through $\TTrap$).
To better understand this evolution, it is convenient to consider the effective mass:
\begin{gather}
	\maThermalgeneral^2 = \frac{1}{a} \pdv{\VThermalgeneral}{a} \Big\vert_{\rm min} \, .
\end{gather}
This effective mass allows us to understand the cosmological evolution by comparing 
it
with $m_a(T)$ and $H(T)$. In figure \ref{fig:evolutionPlotCombined}, we can see how the effective masses (dashed lines) 
dominate over $m_a(T)$ until $T\sim 950 $ MeV. Furthermore, both effective masses dominate over $H(T)$ before trapping. 
While the former is true by construction, the second feature is a condition which is necessary for the self-consistency of the scenario.
At the very least, it is necessary that the effective masses dominate Hubble immediately prior to the transition at $\TTrap$.
Saturating the inequality $\maThermalgeneral(\Tosc) > H(\Tosc)$, we find that it is satisfied in the same regime in which the standard misalignment mechanism underproduces DM. This is because $\maThermalgeneral(\TTrap) \sim H(\TTrap)$ implies $m_a(\TTrap) \sim H(\TTrap)$.
This means that Hubble is subdominant to $m_a(\TTrap)$ and $\maThermalgeneral(\TTrap)$ in the mass range 
\begin{gather}
	m_a \gg 10^{-5} \text{ eV} \, . \label{eq:maMin, strict}
\end{gather}
In this regime, the thermal potential can dominate at least for a short time before $\TTrap$, which is required by the consistency of the trapping scenario. If $q=4$ then $\maThermalgeneral \propto T^2 $ and the mass remains dominant with respect to $H(T)\propto T^2$ for all $T \gtrsim \TTrap$. However, if $q < 4$, then $H(T)$ can dominate at higher temperatures, and we must ask whether the field has sufficient time to settle down to the minimum of the PQ-breaking potential. The numerical results presented in Ref.~\cite{Zhang:2023gfu} suggest that a short duration of oscillations prior to $\TTrap$ is sufficient for $\theta$ to relax to the false minimum, meaning that even for $q<4$ eq.~\eqref{eq:maMin, strict} could be taken as a reasonable range of validity of the trapping scenario, although a dedicated numerical analysis is necessary to confirm that exact range of validity.

For PQ-breaking operators 
involving SM fermions, which feature thermal potentials with $q=2$ (see section \ref{sec:thpotOSM}), we can nonetheless derive a condition which guarantees Hubble domination at $T\gg \TTrap$.
This is possible because the SM fermions become massless at temperatures above weak phase transition, which takes place around $T\approx v = 246$ GeV. 
Above this scale, the temperature dependence transitions from $\VThermalfermion \propto m_f^2T^2$ to $\VThermalfermion \propto T^4$.
Thus, if $\maThermalFermions(T) > H(T)$ at $T\sim v$, then the PQ-breaking potential will continue to dominate at arbitrarily higher temperatures.
We find that this effect guarantees the validity of the trapping scenario for
\begin{gather}
	m_a \gg 10^{-3} \text{ eV}. \label{eq:maMin, loose}
\end{gather}
It is important to emphasize that, although this condition is sufficient to ensure the relaxation of $\theta$ to the minimum of $\VThermalgeneral$, it is not a necessary condition. 
In fact, given the numerical results of \cite{Zhang:2023gfu} mentioned above, the actual range of validity of the trapping scenario might lie much closer to the necessary but not sufficient condition of eq.~\eqref{eq:maMin, strict}. 
As a guide to the reader, we highlight the features ensured by eqs. \eqref{eq:maMin, strict} and \eqref{eq:maMin, loose} on fig. \ref{fig:evolutionPlotCombined}. In subsequent plots, we will show the DM solution only when the condition \eqref{eq:maMin, strict} is satisfied, and for $q=2$ scenarios we will use a dashed line for the mass range in which the condition \eqref{eq:maMin, loose} is not satisfied.

\subsection{Thermal potentials from PQ-breaking operators coupled to SM fields}
\label{sec:thpotOSM}

Now that we have a simple and general, solution, we can consider explicit SM operators which realize couplings of the form in eq.~\eqref{eq:phiOSM}. 
From eq.~\eqref{eq:ToscDM} we know that the relevant temperature 
for the misalignment mechanism is $T\sim 1$ GeV, so to generate a thermal potential we must consider couplings to SM fields that are relativistic at such temperatures. This immediately rules out 3rd generation fermions, $W^\pm$, $Z$, and Higgs bosons, as well as the charm quarks. The candidate operators are thus the $u,d,s,e,$ and $\mu$ Yukawa operators as well as the $GG$ and $FF$ operators. 
Here, we consider the fermion Yukawa operators and the $GG$ operator while we leave $FF$ to future work.
Specifically, we consider fermion operators of the form
\begin{align}
\LPQV
\supset e^{i\deltaPQV} \left(\frac{\phi}{\LambdaUV}\right)^{n} \frac{\sqrt{2}m_f}{v} \bar{L}_{f} H f_R+\text{h.c.}\ ,\label{eq:fermion Yukawa}
\end{align}
where $m_f$, $L_f$ and $f_R$ are the Dirac mass, left-handed component and right-handed component of $f = u,d,s,e,\mu$. The $GG$ coupling 
is taken to be of the form
\begin{gather}
\label{eq:PQVGG}
\LPQV
\supset e^{i\deltaPQV} \left(\frac{\phi}{\LambdaUV}\right)^{n} \alphaStrong G^a_{\mu\nu}G^{a\,\mu\nu}+\text{h.c.}\ , 
\end{gather}
These operators were previously studied in Ref.~\cite{Zhang:2023gfu} with the motivation of using thermal potentials to resolve the domain wall problem in post-inflationary axion scenarios.  
For $T\ll v$, which is the temperature range of interest here, the thermal potential for fermions can be obtained from standard one-loop effective potentials techniques~\cite{Zhang:2023gfu,Quiros:1999jp}
\footnote{Note that 
we find an extra $1/2$ factor in eq.~\eqref{eq:VThermalfermion} 
with respect to Ref.~\cite{Zhang:2023gfu}.}
\begin{gather}
\VThermalfermion \approx - \frac{\ncol}{6} \left(\frac{f_a}{\sqrt{2}\LambdaUV}\right)^n m_f^2 T^2  \cos(n\theta + \deltaPQV)\, ,\label{eq:VThermalfermion}
\end{gather}
where $\ncol = 1$ for leptons and 3 for quarks.
Note that eq.~\eqref{eq:VThermalfermion} assumes that quarks can be described perturbatively when considering the interactions 
in eq.~\eqref{eq:fermion Yukawa}. 
Although such a description breaks down at $T\sim \TQCD \approx 150$ MeV, it is expected to be valid 
up to $\mathcal{O}(1)$ corrections
for $T\sim$ GeV.

Gauge boson interactions have been also taken into account in 
Ref.~\cite{Zhang:2023gfu}, 
and in the case of gluons, 
cf.~eq.~\eqref{eq:PQVGG}, 
they lead to the 
following thermal potential arising at two loops (see Supplemental Material in \cite{Zhang:2023gfu})
\begin{gather}
\label{eq:VthGG}
\VThermalGG \approx - 480 \, \pi \alphaStrong^2 \left(\frac{f_a}{\sqrt{2}\LambdaUV}\right)^n T^4 \cos(n\theta + \deltaPQV) \, .
\end{gather}
Since the relevant dynamics for the misalignment mechanism is 
around $T\sim 1$ GeV, 
in the above equation 
we employ $\alphaStrong (\mu = 1 \, \text{GeV}) \approx 0.5$ \cite{Chetyrkin:2000yt,Herren:2017osy}.

In summary, the PQ-breaking operators considered 
in eqs.~\eqref{eq:fermion Yukawa} and \eqref{eq:PQVGG} 
generate thermal potentials of the form of 
eq.~\eqref{eq:Vthermal general}, 
with the 
following parameters:
\begin{align}
	 \text{Fermions:}&& \lambdaPQV^4 &= \frac{\ncol}{6}m_f^2 \, ,  & q&=2 \, ,  \\
	\text{Gluons:}&& \lambdaPQV^4 &= 480 \, \pi \alphaStrong^2 \, , & q&=4 \, .  
\end{align}
The value of 
$\LambdaUV$ which reproduces the observed 
DM abundance can then be found by plugging the above parameters into eq.~\eqref{eq:LambdaUV DM sol, general}.

In this work, we take a bottom-up approach and study the phenomenological impact of the above operators. However, we recognize that there might be model-building challenges associated with their realization, which we discuss in appendix \ref{app:model-building}.

\subsection{Induced axion VEV} 
\label{sec:axionVEV}
Because the PQ-breaking operators in eqs.~\eqref{eq:fermion Yukawa}-\eqref{eq:PQVGG} violate CP in the presence of a non-zero phase $\deltaPQV$ they also drive a non-zero axion VEV.
This can either be generated radiatively or via long-distance QCD effects. We here discuss which of these effects dominate the 
axion VEV for each of the operators in eqs.~\eqref{eq:fermion Yukawa} and \eqref{eq:PQVGG}.

Consider first the fermion Yukawa operator in eq.~\eqref{eq:fermion Yukawa}. As discussed in Refs.~\cite{Zhang:2022ykd,Zhang:2023gfu}, 
this operator is responsible for a radiatively induced axion VEV.  
Here, we follow the approach of Ref.~\cite{DiLuzio:2024ctr} 
to compute the axion tadpole, $\VPQV = - \sigma a + \ldots$ at one loop. 
After setting $\langle H \rangle = ( 0 \ \ v/\sqrt{2})^T$, 
in the leading-log approximation we find 
\beq 
\sigma (\mu) = - \frac{n\, \ncol}{2\pi^2} 
\left(\frac{f_a}{\sqrt{2}\LambdaUV}\right)^{\!n} 
\frac{m_f^4}{f_a} \sin\deltaPQV \ln \( \frac{v}{\mu} \) ,
\eeq
where $\mu$ denotes the renormalization scale. 
Including the 
contribution of the QCD axion potential, 
we obtain the induced axion VEV
\begin{align}
\label{eq:thetaVEV}
\theta_{\rm eff} &\approx  
 - \frac{n\, \ncol}{2\pi^2} 
 \left(\frac{f_a}{\sqrt{2}\LambdaUV}\right)^{\!n} 
 \!\frac{m_f^4}{\chiQCD}
 \sin\deltaPQV 
 \,\ln\left(\frac{v}{1 \, \text{GeV}}\right) 
 \, ,  
\end{align}
where we took $\mu = 1$ GeV 
to assess the 
nEDM bound. 
This VEV is generated by all the fermion Yukawa operators in eq.~\eqref{eq:fermion Yukawa}, but it is particularly relevant for leptons.

In the case of quarks, the operator in 
eq.~\eqref{eq:fermion Yukawa} also yields a long-distance  
QCD contribution to the axion potential  
\begin{align}
V^{qq}_{\text{long-dis.}} \approx -\left(\frac{f_a}{\sqrt{2}\LambdaUV}\right)^n m_q 
\, \langle \overline{q} q \rangle \cos(n\theta+\deltaPQV) \, , 
\end{align}
where $\langle \overline{q} q \rangle \approx (240 \, \text{MeV})^3$,  
for $q=u,d,s$, 
is the quark condensate evaluated at the renormalization scale 
of 1 GeV. Upon minimizing $ V_{\rm QCD} + V^{qq}_{\text{long-dis.}}$, the induced axion VEV is  
\begin{gather}
\label{eq:thetaVEVLRqq}
	\thetaVEV \approx - n
 \left(\frac{f_a}{\sqrt{2}\LambdaUV}\right)^n \frac{m_q \langle \overline{q} q \rangle}{\chiQCD} 
 \sin\deltaPQV \, .
\end{gather}
Note that in the case of quarks the VEV in 
eq.~\eqref{eq:thetaVEVLRqq} dominates, 
by several orders of magnitude, 
with respect to the radiative one 
in eq.~\eqref{eq:thetaVEV}.
Therefore, eq.~\eqref{eq:thetaVEVLRqq} is relevant for quark operators 
while eq.~\eqref{eq:thetaVEV} yields the dominant contribution in the case of leptons. 

Next, we consider the axion potential induced by the 
axion coupling to gluons, due to the PQ-breaking operator in 
eq.~\eqref{eq:PQVGG}. In this case, the axion potential 
is dominated by the long-distance QCD contribution, which reads 
\begin{align}
V^{GG}_{\text{long-dis.}} \approx - \left(\frac{f_a}{\sqrt{2}\LambdaUV}\right)^n 2 \pi
\, \expval{\frac{\alphaStrong}{\pi} G G} 
\cos(n\theta+\deltaPQV) \, , 
\end{align}
where the gluon condensate is 
$\langle \frac{\alphaStrong}{\pi} G G \rangle \approx 
\left(330 \text{ MeV}\right)^4$ \cite{Shifman:1978bx}. 
Upon minimizing $V_{\rm QCD} + V^{GG}_{\text{long-dis.}}$, 
one obtains the axion VEV \cite{Zhang:2022ykd}
\begin{equation}
    \thetaVEV \approx 
    - 2 \pi n 
    \frac{\expval{{\frac{\alphaStrong}{\pi} GG}}}{\chiQCD} \left(\frac{f_a}{\sqrt{2}\LambdaUV}\right)^n \sin \deltaPQV
    \, .  \label{eq:thetaVEVGG}
\end{equation}

\subsection{Direct and indirect scalar 
axion
interactions}
\label{sec:scalar interactions}
The PQ-breaking operators we are considering here lead to scalar interaction terms of the form
\begin{gather}
	\mathcal{L}\supset g_{a\overline{f}f} a \overline{f}f \, . \label{eq:expanded coupling}
\end{gather}
In our framework, there are two ways such an interaction can arise. The first is directly from 
the fermion interaction in eq.~\eqref{eq:fermion Yukawa}, which 
contains a term of the form of eq.~\eqref{eq:expanded coupling} with
\begin{gather}
g_{a\overline{f}f}^{\text{direct}(\overline{f}f)} = n\frac{m_f}{f_a}\left(\frac{f_a}{\sqrt{2}\LambdaUV}\right)^n\sin\deltaPQV \, . \label{eq:expanded coupling - fermion direct contribution}
\end{gather} 
This coupling arises both from the lepton and the quark Yukawa operators. At low energies, through the matrix element $\bra{N}m_q \overline{q}q\ket{q}=f_q m_N \bar{\psi}_N \psi_N$ \cite{Zhang:2023gfu,Cline:2013gha} the latter operator gives rise to a scalar axion-nucleon interaction of the form
\begin{gather}
g_{a\overline{N}N}^{\text{direct}(\overline{q}q)} = n\frac{f_q m_N}{f_a}\left(\frac{f_a}{\sqrt{2}\LambdaUV}\right)^n\sin\deltaPQV \, ,\label{eq:expanded coupling - quark direct contribution}
\end{gather}
where we use the values $f_q$ provided in \cite{Cline:2013gha}.
For the $GG$ coupling, a direct contribution to the axion-nucleon coupling is also generated by the interaction with the gluonic component of the nucleons. To evaluate this, we use the matrix element $ m_N \bar{\psi}_N \psi_N \approx -\frac{9}{8\pi} \bra{N}\alphaStrong GG \ket{N} $ \cite{Shifman:1978bx} to identify the coupling
\begin{gather}
 g_{a\overline{N}N}^{\text{direct}(GG)}  = n \frac{16\pi}{9} \frac{m_N}{f_a} \left(\frac{f_a}{\sqrt{2}\LambdaUV}\right)^n \sin\deltaPQV 
 \, .
\end{gather}
Alternatively, a scalar axion-nucleon coupling can also arise in an indirect way through the axion VEV. In particular, if the axion has a VEV $\thetaVEV \neq 0$, then the latter generates a scalar axion-nucleon interaction 
given by 
the standard isospin-symmetric formula of Moody and Wilczek~\cite{Moody:1984ba},  
with the correct extra 1/2 factor~\cite{Bertolini:2020hjc}
\begin{gather}
g_{a\overline{N}N}^{\text{indirect}(\thetaVEV)}  = \frac{\thetaVEV}{f_a} \frac{m_u m_d}{m_u + m_d} \frac{\bra{N}\bar{u}u+\bar{d}d\ket{N}}{2}\approx \thetaVEV \frac{f_s}{f_a} 
\, ,  
\label{eq:indirect axion-nucleon coupling}
\end{gather}
where $f_s \approx 13\text{ MeV}$ and we ignored isospin breaking effects which lead to corrections at the 10\% level \cite{Bertolini:2020hjc}. To evaluate this indirect contribution, we plug in the relevant $\thetaVEV$ contribution determined 
in section \ref{sec:axionVEV} 
for the interactions in eq.~\eqref{eq:fermion Yukawa} and \eqref{eq:PQVGG}.

For quark-Yukawa and $GG$ operators, both of which generate direct axion-nucleon couplings, the direct and indirect contributions are comparable, although the indirect contribution is slightly dominant. 
Lepton-Yukawa operators do not generate a direct axion-nucleon interaction. Nevertheless, as we will see shortly, all of the above scalar interactions give rise to long-range forces. 
Therefore, even though $g_{a\overline{e}e}^{\rm direct}$ and $g_{a\overline{N}N}^{\rm indirect}$ naively correspond to different observables, it will become useful to compare the amplitude of the two couplings:
\begin{gather}
    \frac{g_{a\overline{e}e}^{\text{direct}(\overline{e}e)}}{g_{a\overline{N}N}^{\text{indirect}(\thetaVEV)} } \approx \frac{2\pi^2}{\ln(v/\mu)}\frac{\chiQCD}{f_s m_e^3} \approx 7 \times 10^7. \label{eq:direct vs indirect for leptons}
\end{gather}
For leptons, we will see that this implies a stronger constraint from the direct axion-lepton coupling.

\subsection{Constraints}\label{sec:constraints}
Now that we have a set of candidate PQ-breaking operators, we know which parameters reproduce the DM 
relic density, and we know the implied axion VEVs and couplings, 
we will discuss 
constraints on the parameter space of those PQ-breaking scenarios. 
We will consider constraints stemming from the EDM of the neutron, 
CP-violating axion interactions 
in the form of 
monopole-monopole and monopole-dipole forces
(for a recent review, see \cite{DiLuzio:2023lmd}), 
as well as astrophysical bounds. 

\subsubsection{nEDM}
\label{sec:nEDM constraints}

As discussed above, the nEDM constrains the axion VEV to $|\thetaVEV| \lesssim 10^{-10}$.\footnote{Other 
contributions to EDMs
stemming from direct axion exchange, 
recently computed in Refs.~\cite{DiLuzio:2020oah,DiLuzio:2023cuk}, 
turn out to yield subleading constraints on the interactions 
of eqs.~\eqref{eq:fermion Yukawa} and \eqref{eq:PQVGG}, 
since they are associated to 
higher-dimensional CP-violating operators 
featuring an extra $1/f_a$ suppression.} Applying this condition to 
eq.~\eqref{eq:thetaVEV} for the lepton Yukawa operators, 
eq.~\eqref{eq:thetaVEVLRqq} for quark Yukawa operators, and to eq.~\eqref{eq:thetaVEVGG} for the gluon operator we can derive upper limits on the scale $\LambdaUV$.

For the quark Yukawa, once we assume the value of $\LambdaUV$ required for trapped misalignment to account for DM, 
eq.~\eqref{eq:LambdaUV DM sol, general}, then the axion VEV generated by the quark condensate is large enough to firmly rule out the scenario:
\begin{gather}
\abs{\thetaVEV} \approx 3.2 \times 10^{-7} \left( \frac{\dmfraction}{1.0} \right)^{\frac{2(\beta+1)}{\beta+3}} \left(\frac{m_{s}}{m_q}\right)\frac{\sin \deltaPQV}{\abs{\thetaTrap}^{\frac{4 (\beta+1)}{\beta+3}}} \, ,
\end{gather}
where we normalized the quark mass $m_q$ to the strange quark mass $m_s$ because this yields the smallest $\thetaVEV$ for the operators considered here. It is evident that a coupling to either of the quarks cannot be viable because it predicts an nEDM larger than the observational bound.

On the other hand, the leptons receive only the radiative contribution to $\thetaVEV$. Therefore, the value of $\thetaVEV$ implied by the DM solution is much smaller 
and can be compatible with the nEDM bound:  
\begin{gather}
\abs{\thetaVEV} \approx  2.4\times 10^{-13} \left( \frac{\dmfraction}{1.0} \right)^{\frac{2(\beta+1)}{\beta+3}} \left(\frac{m_{\ell}}{m_e}\right)^2\frac{\sin \deltaPQV}{\abs{\thetaTrap}^{\frac{4 (\beta+1)}{\beta+3}}}  \, ,
\end{gather}
where $m_\ell$ (with $\ell=e,\mu$) is the lepton mass. Since $\thetaVEV$ scales with $m_\ell^2$, only scenarios with an electron coupling are compatible with nEDM constraints. If we consider 
for example
an axion coupling to the muon Yukawa then the enhancement $(m_\mu/m_e)^2\approx 4 \times 10^4$ pushes well $\thetaVEV$ above $10^{-10}$. 

Interestingly, if the axion is coupled to $GG$, then the predicted values of $\thetaVEV$ are very close to the observational bound:
\begin{gather}
\abs{\thetaVEV} \approx  1.2\times 10^{-10} \left( \frac{\dmfraction}{1.0} \right)^{\frac{2 (\beta+2)}{\beta+3}} \frac{\sin\deltaPQV}{\abs{\thetaTrap}^{\frac{4 (\beta+2)}{\beta+3} }}   \, . \label{eq:thetaVEVGG assuming DM}
\end{gather}
If we assume that the axion makes up all of the DM, i.e.~$\dmfraction=1$, then for the gluon scenario the compatibility with the nEDM is controlled by the last term in eq.~\eqref{eq:thetaVEVGG assuming DM}. This is expected to be $\mathcal{O}(1)$ in the absence of tuning, which means that while the gluon scenario is compatible with nEDM in an $\mathcal{O}(1)$ fraction of the parameter space (in $\thetaini$ and $\deltaPQV)$, it generically predicts an observable nEDM just beyond the current bound.

Since for the gluon operator, the axion-nucleon interaction is dominantly generated by the axion VEV, this corresponds to the usual picture in which the nEDM is the most important constraint.
For the lepton Yukawa, we have seen that the scalar interactions generated directly by eq.~\eqref{eq:fermion Yukawa} dominate by far the 
one due to 
contribution from the axion VEV, see eq.~\eqref{eq:direct vs indirect for leptons}. This means that the nEDM constraint on the axion VEV is not necessarily the dominant constraint for 
lepton-Yukawa PQ-breaking operators.

\subsubsection{Monopole-monopole forces}
Since in the lepton-Yukawa 
PQ-breaking scenario we have scalar interactions which are not derived from the axion VEV, we have the possibility that constraints from long-range forces can become important. In particular, 
the scalar interactions discussed in section \ref{sec:scalar interactions} generate a Yukawa potential of monopole-monopole type,
\begin{gather}
	V_{\rm Yukawa} = -  g_{a\overline{f}f}^2 \frac{e^{-m_a r}}{4\pi r} \, ,
\end{gather}
where $r$ is the distance between two bodies. Such a potential gives rise to long-range forces constrained by searches for fifth forces 
and for equivalence principle (EP) violation. 
Fifth force searches constrain violations of the inverse-square law through the standardized parameter $ \alpha $ defined via \cite{Adelberger:2009zz}:
\begin{gather}
	V_{\rm tot,12} = V_{G,12} (1+\alpha e^{-m_a r}) \, ,
\end{gather}
where $ V_{\rm tot ,12} $ and $ V_{G,12} $ are the total potential energy and the gravitational potential energy, respectively, between two test masses labelled 1 and 2. We apply the constraints on $\alpha$ which are collected in \cite{Lee:2020zjt}.\footnote{We note that Refs.~\cite{Zhang:2022ykd,Zhang:2023gfu} appear to have used the older bounds reported in \cite{Adelberger:2009zz}.  Those older bounds are significantly more permissive 
for axion masses in the $10^{-2}$ eV to $10^{-1}$ eV range, 
where we find the fifth force constraints to play an important role.}

The exact form of $\alpha$ depends on which fermion we couple to 
and on the material composition of the test masses.
If the coupling is with electrons, and the test masses have each $N$ atoms consisting of $Z$ electrons and $A$ nucleons, then the 
axion-mediated potential is
\begin{gather}
	V_{\rm Yukawa,12} = - (N_1Z_1) (N_2Z_2)   \left(g_{a\overline{e}e}^{\text{direct}(\overline{e}e)}\right)^2 \frac{e^{-m_a r}}{4\pi r} \, ,  \label{eq:electron long-range Yukawa potential}
\end{gather}
which is compared to the gravitational potential 
\begin{gather}
	V_{G,12} = - (N_1 A_1) (N_2 A_2) \frac{u^2G}{r} \, ,
\end{gather}
where $ u = 931.5 $ MeV is the atomic mass unit and $G  = (8\pi \mplanck^2)^{-1}$ is the gravitational constant.
For the axion-electron coupling, the parameter $\alpha$ then takes the form
\begin{gather}
	\alpha_{a\overline{e}e} = \frac{V_{\rm Yukawa,12}}{V_{G,12}} e^{m_a r} = \frac{Z_1 Z_2}{A_1 A_2} \frac{\left(g_{a\overline{e}e}^{\text{direct}(\overline{e}e)}\right)^2}{4\pi u^2 G} = \frac{Z_1 Z_2}{A_1 A_2} \frac{m_e^2}{f_a^2}\left(\frac{f_a}{\sqrt{2}\LambdaUV}\right)^{2n}\frac{n^2\sin^2\deltaPQV}{4\pi u^2G}\,,\label{eq:FF constraint electron}
\end{gather}
in agreement with \cite{Zhang:2022ykd}.
Since the most relevant bounds on $ \alpha $ are set from the Eöt-Wash experiment~\cite{Lee:2020zjt} employing Aluminum ($ Z/A \sim 0.482 $) and Copper $ (Z/A\sim 0.456) $ for test mass 1 and 2,
respectively, we use these numbers in our constraints. 

Note that the indirect axion-nucleon coupling \eqref{eq:indirect axion-nucleon coupling} yields a Yukawa potential similar to 
eq.~\eqref{eq:electron long-range Yukawa potential} except with $g_{a\overline{e}e}^{\text{direct}(\overline{e}e)} \to g_{a\overline{N}N}^{\text{indirect}(\thetaVEV)}$ and $Z_i\to A_i$. For the electron, the comparison in eq.~\eqref{eq:direct vs indirect for leptons} makes it clear that the direct interaction is significantly more constraining, such that eq.~\eqref{eq:FF constraint electron} is the relevant parameter to be considered.

If we consider instead a Yukawa coupling to a quark $q$ or  the gluons, then a similar bound arises dominantly from the indirect axion-nucleon coupling.
The $\alpha$ parameter is found from eq.~\eqref{eq:FF constraint electron} with the modifications $g_{a\overline{e}e}^{\text{direct}(\overline{e}e)} \to g_{a\overline{N}N}^{\text{indirect}(\thetaVEV)}$ and $Z_i\to A_i$. Plugging in the $\thetaVEV$ generated by the quark Yukawa, eq.~\eqref{eq:thetaVEVLRqq}, 
the parameter $\alpha$ takes the form
\begin{gather}
	\alpha_{a \overline{q}q}  = \frac{m_q^2 \expval{\overline{q}q}^2}{\chiQCD^2} \frac{f_s^2}{f_a^2}  \left(\frac{f_a}{\sqrt{2}\LambdaUV}\right)^{2n}\frac{n^2\sin^2\deltaPQV}{4\pi u^2G}\, .\label{eq:FF constraint quark}
\end{gather}
Plugging in the $\thetaVEV$ generated by a gluon coupling, eq.~\eqref{eq:thetaVEVGG}, we obtain 
\begin{gather}
	\alpha_{a GG}  = \frac{\pi^2\expval{{\frac{\alphaStrong}{\pi} GG}}^2}{\chiQCD^2} \frac{f_s^2}{f_a^2}  \left(\frac{f_a}{\sqrt{2}\LambdaUV}\right)^{2n}\frac{n^2\sin^2\deltaPQV}{4\pi u^2G}\, .\label{eq:FF constraint gluon}
\end{gather}
In both of these cases, the atomic composition of the test masses cancels out because both gravity and the fifth force couple dominantly to the nucleons. 

Searches for violation of the EP are very similar to fifth force constraints, in the sense that they also constrain long-range forces. However, while fifth force searches look for any violation of the 
inverse-square law of gravity, searches for EP violation are sensitive to long-range forces that are material-dependent. Generally, EP-violating forces also violate the inverse-square law (fifth force), but not all such ``fifth forces'' violate EP. In our case, not only does the Yukawa interaction deviate from the inverse-square law, but it is also material-dependent because it describes an interaction with only a single SM field, so our scenario is subject to both types of constraints.
EP constraints are parameterized slightly differently from fifth force constraints~\cite{Adelberger:2009zz}: 
\begin{gather}
	V_{\rm tot,12} = V_{G,12} \left(1 + \tilde{\alpha} \left[\frac{\tilde{q}_1}{g_{a\overline{f}f} N_1A_1}\right] \left[\frac{\tilde{q}_2}{g_{a\overline{f}f}N_2A_2}\right] e^{-m_a r}\right) \, , 
\end{gather}
where $\tilde{q}_i$ are the total Yukawa charges. In the case of the $a\overline{e}e$ coupling, the total charge is $ \tilde{q}_i = g_{a\overline{e}e} N_i Z_i $. The prefactor is thus exactly the $ (Z/A)^2 $ factor of $ \alpha $, such that the EP parameter for the electron coupling is 
\begin{gather}
	\tilde{\alpha}_{a\overline{e}e} = \frac{m_e^2}{f_a^2}\left(\frac{f_a}{\sqrt{2}\Lambda}\right)^{2n}\frac{n^2\sin^2\delta}{4\pi u^2G}\,,\label{eq:EP constraint}
\end{gather}
which is in agreement with Ref.~\cite{Zhang:2022ykd}. For the case of the quark and gluon couplings, the total charge is set by the coupling to nucleons, $ \tilde{q}_i = g_{a\overline{N}N}^{\text{indirect}(\overline{e}e)} N_i A_i $, so that 
the $\alpha$ and $\tilde{\alpha}$ parameters coincide:
\begin{gather}
    \tilde{\alpha}_{a\overline{q}q} = \alpha_{a\overline{q}q} \qq{and}  \tilde{\alpha}_{aGG} = \alpha_{aGG} \, .
\end{gather}
Limits on $ \tilde{\alpha} $ are reported in a coupling-dependent manner. Ref.~\cite{Adelberger:2009zz} reports limits on $ \tilde{\alpha} $ arising from either couplings to $Z$ (i.e.~electrons or protons), $B$ (baryon number, here equivalent to $A$) or to the particular combination $B-L$ (baryon minus lepton number). In our scenario, the $Z$-coupling bound is relevant for the electron coupling and the $B$-coupling bound is relevant for quark couplings. 
The most up-to-date constraints on EP violating couplings to $B$ are, in descending order of axion mass, Eöt-Wash 1999 \cite{Smith:1999cr}, Eöt-Wash 2008 \cite{Schlamminger:2007ht} and MICROSCOPE \cite{Berge:2017ovy}. MICROSCOPE is dominant for 
$m_a \lesssim 10^{-12}$ eV, therefore the older Eöt-Wash constraints collected in \cite{Adelberger:2009zz} are more relevant for our scenario.\footnote{Refs.~\cite{Zhang:2022ykd,Zhang:2023gfu} appear to 
e imposing constraints from a coupling to $B-L$, which leads to an $\mathcal{O}(1)$ 
difference relative to the $\tilde{\alpha}$ constraint from a coupling to $B$.}

\subsubsection{Monopole-dipole forces}
The long-range forces described in the previous section involve purely scalar interactions (monopole-monopole). 
Alternatively, one of the vertices can be exchanged for the usual 
axion
pseudo-scalar coupling, which leads to monopole-dipole 
forces. Typically (although not in our PQ-breaking electron-Yukawa scenario), monopole-monopole forces are penalized by $\thetaVEV^2$, which severely limits 
their
experimental reach. Search for monopole-dipole interactions are motivated by reducing this to a linear $\thetaVEV$ suppression. The trade-off is that 
pseudo-scalar 
interactions are spin-suppressed and, in the case of  
dipole-dipole forces involving only pseudo-scalar couplings, also
limited by a large background from ordinary magnetic forces, which restricts their sensitivity. Monopole-dipole forces are often seen as an optimal compromise and the ARIADNE experiment \cite{Arvanitaki:2014dfa,ARIADNE:2017tdd} 
 could place a stricter bound on $\thetaVEV$ than nEDM experiments in the 
axion mass
range in which it is sensitive. In our case, the PQ-breaking electron-Yukawa scenario predicts a scalar interaction much stronger than the indirect contribution from $\thetaVEV$, which suggests that monopole-dipole searches 
could play an important role \cite{DiLuzio:2024ctr}.  

ARIADNE consists of a mobile tungsten source mass, which generates a potential through the scalar interaction, and a helium-3 target mass which responds through the pseudo-scalar interaction. The projections are given in terms of axion-nucleon couplings. Since we are primarily interested in the axion-electron scalar coupling, and since we do not expect it to matter which part of the atom the axion potential is generated from, we rescale the projections 
using
\begin{gather}
g_{a\overline{N}N}^{\rm effective} = \frac{Z}{A}g_{a\overline{e}e}^{\text{direct}(\overline{e}e)} \, ,
\end{gather}
where $Z = 74$ and $A=183.84$ 
are the average numbers of electrons and nucleons in each tungsten atom. Using this effective interaction for the electron scenario, and the $g_{a\overline{N}N}^{\rm indirect}$ for the gluon scenario, we present projections 
for ARIADNE in figure \ref{fig:summaryPlots}, 
corresponding to the initial experimental phase. 
Currently, the strictest 
bounds on monopole-dipole interactions arise from a combination of laboratory monopole-monopole searches and astrophysical limits 
on pseudo-scalar couplings \cite{Raffelt:2012sp,OHare:2020wah}. We also include this constraint in figure \ref{fig:summaryPlots}, although it is generally subdominant to monopole-monopole constraints.

Several other experiments have established bounds or are attempting to improve the sensitivity 
of monopole-dipole searches on
other couplings 
\cite{Wineland:1991zz,Heckel:2008hw,Hoedl:2011zz,Terrano:2015sna,Crescini:2016lwj,Crescini:2017uxs,Stadnik:2017hpa,Lee:2018vaq,Dzuba:2018anu,Crescini:2020ykp,Fan:2023hci,Agrawal:2023lmw,Baruch:2024frj,Baruch:2024fbh,Wei:2022ggs,Agrawal:2022wjm}, but as they do not improve upon the constraints or projections presented here we do not include them in our plots.

\subsubsection{Astrophysical axion bounds}

In a given QCD axion model,  
the astrophysical constraints on the standard axion couplings to photons, electrons and nucleons 
can be translated into an upper bound on the axion mass.
Standard KSVZ \cite{Kim:1979if,Shifman:1979if} and 
DFSZ \cite{Zhitnitsky:1980tq,Dine:1981rt}
scenarios 
require $m_a \lesssim 10^{-2} $ eV, 
from the limits on the axion couplings 
to nucleons and electrons, 
stemming respectively 
from the neutrino burst duration of the SN1987A 
and the white-dwarf cooling rates / red giants evolution in globular clusters (for a collection of bounds, see e.g.~\cite{DiLuzio:2020wdo,DiLuzio:2021ysg}). On the other hand, 
so-called astrophobic axion models 
\cite{DiLuzio:2017ogq} 
(see also \cite{Bjorkeroth:2018ipq,Bjorkeroth:2019jtx,DiLuzio:2022tyc,Badziak:2023fsc,Takahashi:2023vhv,Badziak:2024szg}) based on non-universal DFSZ variants, 
allow us to
relax the above constraints by about one order of magnitude. 
Those models simultaneously suppress the 
axion couplings to nucleons and electrons, at the cost of a 
single tuning of $\mathcal{O}(10\%)$. 
In the following, we consider astrophysical constraints on the axion mass  
for a few representative axion models, including~\cite{DiLuzio:2017ogq}
\begin{align}
        m_a &\lesssim 8\times 10^{-3} \text{ eV} 
        &&\text{KSVZ with }E/N = 0 \, , \\
		m_a &\lesssim 1.5 \times 10^{-2}\text{ eV} 
        &&\text{DFSZ with }E/N = 8/3 \, , \\
		m_a &\lesssim 2\times 10^{-1}\text{ eV} 
        &&\text{Astrophobic M2 with }E/N = 8/3 \, , 
\end{align}
where $E/N$ denotes the ratio of the QED to QCD anomalies with the PQ current, entering the axion coupling to photons. The above limits 
are reported in fig.~\ref{fig:summaryPlots} as vertical 
green lines. 

Moreover, the axion scalar coupling in eq.~\eqref{eq:expanded coupling} induces additional cooling channels in stars. Bounds on such interactions from resonant axion production in the core of red giants
were derived in Ref.~\cite{Hardy:2016kme} which provides a limit 
on the scalar axion-electron coupling of
\begin{gather}
	\alpha_{a\overline{e}e} \lesssim 4\times 10^{-32} = \alpha_{a\overline{e}e}^{\rm max} \, ,
\end{gather}
where $\alpha_{a\overline{e}e}\equiv \left(g_{a\overline{e}e}^{\text{direct}(\overline{e}e)}\right)^2 / (4\pi)$. 
This implies a lower bound on $\LambdaUV$,  
given by 
\begin{gather}
	\LambdaUV \gtrsim  \frac{f_a}{\sqrt{2}}\left(\frac{n m_e \abs{\sin(\deltaPQV)}}{f_a \sqrt{4\pi \alpha_{a\overline{e}e}^{\rm max}}}\right)^{1/n} \, .
\end{gather}
Ref.~\cite{Hardy:2016kme} also sets a limit on the scalar 
axion-nucleon coupling:
\begin{gather}
	\alpha_{a\overline{N}N} \lesssim 1\times 10^{-25} = \alpha_{a\overline{N}N}^{\rm max} \, , \label{eq:cooling on electron}
\end{gather}
where $\alpha_{a\overline{N}N} = \left(g_{a\overline{N}N}^{\text{indirect}(\thetaVEV)}\right)^2 / (4\pi)$. In the axion-gluon scenario, this implies an upper limit of 
\begin{gather} 
	\LambdaUV \gtrsim  \frac{f_a}{\sqrt{2}}\left(\frac{\sqrt{\pi} n f_s \expval{{\frac{\alphaStrong}{\pi} GG}}  \abs{\sin(\deltaPQV)}}{f_a \chiQCD \sqrt{ \alpha_{a\overline{N}N}^{\rm max}}}\right)^{1/n} \, . \label{eq:cooling on GG}
\end{gather}
A similar constraint applies for the quarks which we do not report here because the axion-quark scenario is already ruled out by nEDM constraints.
Note that the astrophysical constraints on $\LambdaUV$ from stellar cooling, eqs.~\eqref{eq:cooling on electron} and \eqref{eq:cooling on GG}, are weaker than the constraints set by long-range forces and nEDM. Therefore, these constraints play no major role in the phenomenology studied here.

\subsection{Experimental prospects and testability}
We have evaluated each of the above scenarios, 
implementing a PQ-breaking axion-SM coupling either through a lepton Yukawa, a quark Yukawa or through $GG$, by comparing the DM solution eq.~\eqref{eq:LambdaUV DM sol, general} to the constraints of section \ref{sec:constraints}.
Starting with nEDM constraints, see section \ref{sec:nEDM constraints}, we can immediately narrow down the list of candidate operators to only the axion-electron scenario and the axion-gluon one. Both of these scenarios exhibit interesting phenomenology. 

\begin{figure}
    \centering
    \includegraphics[width=0.7\textwidth]{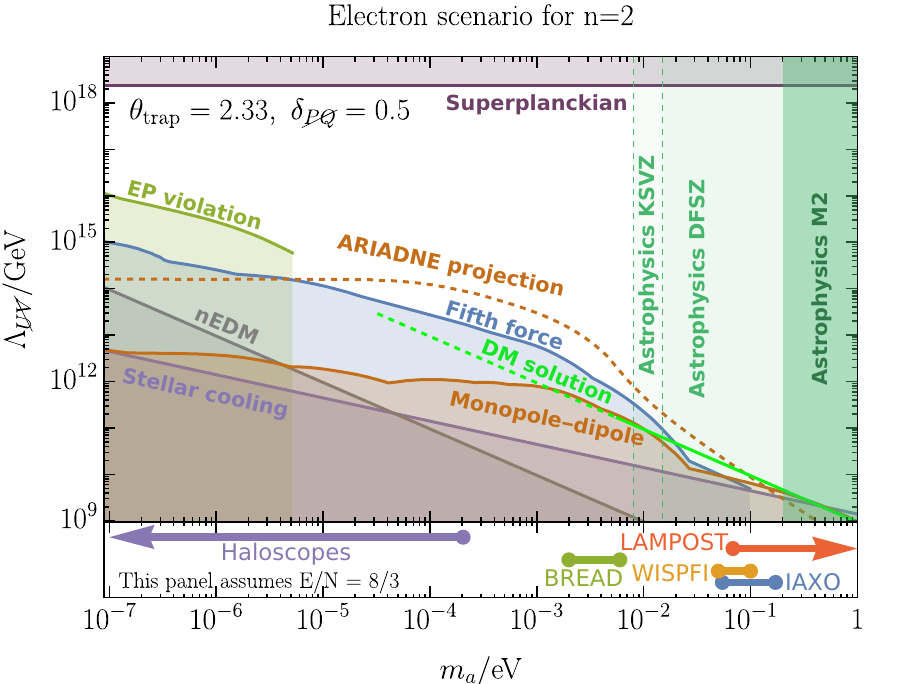} \\
    \vspace{0.5cm}
    \includegraphics[width=0.7\textwidth]{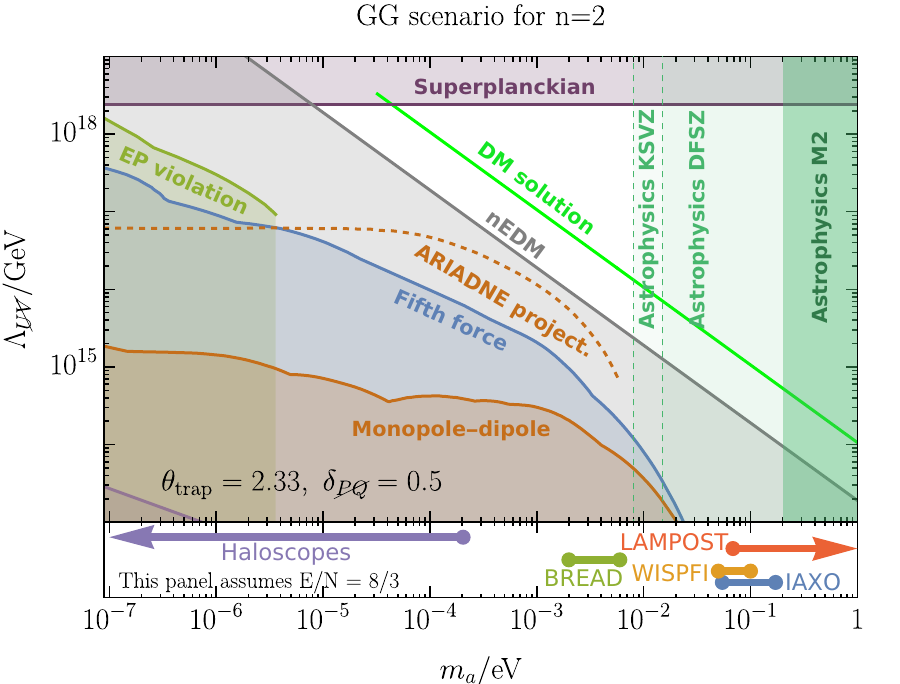}
    \caption{Summary plots for the $(\phi/\LambdaUV)^n$ coupling to 
    the electron-Yukawa (top panel) 
    and $GG$ operator (bottom panel). The trapped misalignment DM solution eq.~\eqref{eq:LambdaUV DM sol, general} (green line) is compared to the constraints discussed in section \ref{sec:constraints} (shaded regions). The green dar matter solution is dashed in the regime in which the thermal potential does not dominate Hubble at arbitrarily high temperatures, 
    eq.~\eqref{eq:maMin, loose}, and the line is 
    interrupted 
    when the consistency condition in eq.~\eqref{eq:maMin, strict} is violated. 
    The plot assumes $n=2$ and a representative $\mathcal{O}(1)$ choice for the initial angle $\thetaini$ and the misalignment 
    angle $\deltaPQV$, which are not tuned. This result demonstrates that there are viable solutions in which trapped misalignment produces QCD axion DM at 
    $m_a \gg 10^{-5}$ eV. These solutions are testable with a moderate improvement in the sensitivity of 
    nEDM and 
    long-range force searches, 
    as well as axion experiments based on the axion-photon 
    coupling (cf.~lower panels in the figure), 
    but do require in the PQ-breaking electron scenario the assumption of an astrophobic axion model to be compatible with 
    astrophysical constraints.}
    \label{fig:summaryPlots}
\end{figure}

In the axion-electron scenario, we only have a contribution to $\thetaVEV$ at the loop level, which allows us to consider stronger couplings (lower $\LambdaUV$) without violating nEDM constraints. These stronger couplings give rise to a direct scalar interaction, eq.~\eqref{eq:expanded coupling - fermion direct contribution}, which dominates 
with respect to the indirect coupling stemming from $\thetaVEV$. This scalar interaction leads to significant constraints from long-range forces which rule our trapped misalignment solutions for axion DM with $m_a \lesssim 2\times 10^{-2}$ eV. This means that if the QCD axion is implemented as in conventional KSVZ or DFSZ models,  
the astrophysical upper limits on $m_a$ exclude the remaining parameter space. However, in 
astrophobic axion models, the axion-electron 
PQ-breaking scenario features trapped misalignment DM solutions in the mass range $2\times 10^{-2}$ eV $\lesssim m_a \lesssim 2\times 10^{-1}$ eV. This class of solutions is especially interesting because it is testable with a moderate increase in the sensitivity of fifth force constraints or monopole-dipole forces. For long-range forces to be sensitive to this scenario, improvements have to be made on deviations from gravity on length scales of $\sim 5\, \mu$m. Furthermore, assuming a photon coupling for an axion with $E/N = 8/3$, such as DFSZ-I or the 
astrophobic M2 model, this scenario motivates axion masses in reach of the proposed/future experiments 
BREAD \cite{BREAD:2021tpx},
WISPFI \cite{Batllori:2023pap}, LAMPOST \cite{Baryakhtar:2018doz} and IAXO \cite{IAXO:2019mpb}. The scenario is summarized in the top half of figure \ref{fig:summaryPlots}. 

In the axion-gluon scenario, the gluon condensate generates a much larger $\thetaVEV$ for a given coupling compared to the axion-electron scenario. However, because $\VThermalgeneral \propto T^4$ in this case, correspondingly weaker couplings are necessary for the trapped misalignment mechanism to be viable. Interestingly, 
for an $\mathcal{O}(1)$ fraction of the possible misalignment angles $\thetaini$ and $\deltaPQV$ the scenario survives current nEDM constraints. 
Since in this case, the scalar coupling is (barely) dominated by the indirect contribution from $\thetaVEV$, we are in the standard scenario in which the nEDM constraints dominate with respect to all constraints from long-range forces, be it monopole-monopole or monopole-dipole. This means that given suitable $\thetaini$ and $\deltaPQV$, trapped misalignment can account for QCD axion DM for $m_a$ between few $\times 10^{-5}$ eV and the upper bound set by astrophysics. Because $\thetaVEV$ 
is never much smaller than $10^{-10}$, this scenario is testable by a moderate improvement in the nEDM sensitivity. In addition to the above-mentioned experiments, the axion-gluon scenario motivates QCD axion DM in reach of BREAD \cite{BREAD:2021tpx} and a large number of other haloscopes with sensitivities at $m_a \lesssim 10^{-4}$ eV, see e.g.~\cite{AxionLimits} for an overview. This scenario is summarized in the lower half of figure \ref{fig:summaryPlots}.

\section{Conclusions}
\label{sec:concl}

In this work, we have studied how much room there is for PQ-breaking potentials to modify the misalignment 
mechanism for axion DM 
within the constraints set by the nEDM and axion-mediated forces. We have focused on the pre-inflationary scenario in which the axion field is initially homogeneous in space.

In the absence of tuning of the CP-violating phase $\deltaPQV$, sources of PQ breaking which are constant in temperature can at most limit the degree to which DM is under-produced by misalignment at large axion masses, i.e.~$m_a \gtrsim 10^{-3}$ eV, in agreement with the existing literature \cite{Jeong:2022kdr}.
Also, in this regime we do not get a fraction of DM 
that is larger than $\Omega_{a} / \Omega_{\rm DM} \lesssim \text{few} \times 10^{-4}$. 

However, by employing a particular class of 
PQ-breaking operators involving also SM fields \cite{Zhang:2022ykd,Zhang:2023gfu}, we find that the 
extra 
temperature dependence 
of the axion potential 
allows for a significant modification of the axion DM abundance from misalignment.
In particular, the same type of trapped misalignment realized for constant temperature potentials in Ref.~\cite{Jeong:2022kdr}, 
with the temperature-dependent potentials of Refs.~\cite{Zhang:2022ykd,Zhang:2023gfu}, allows for solutions in which the QCD axion makes up all the observed DM at masses much larger than what is conventionally expected, i.e.~in the regime $10^{-5}$ eV $\lesssim m_a \lesssim 10^{-1} $ eV. This is realized by coupling the axion to SM fields through a $\phi^n$ operator with $n\geq 2$. We have studied the scalar Yukawa interactions, axion VEVs and long-range forces implied by such couplings. In this respect, we extended the work of 
Refs.~\cite{Zhang:2022ykd,Zhang:2023gfu} by considering additional contributions to the axion VEV, taking the indirect contribution to the scalar interactions into account, updating bounds from searches for fifth forces and 
EP violation, and by extending 
them to both monopole-monopole and monopole-dipole forces. We find that, among the PQ-breaking terms coupled 
to SM operators, the electron Yukawa and the $GG$ terms allow for a viable phenomenology while sizeably 
affecting the production of axion DM from misalignment.  SM particles heavier than $\mathcal{O}$(GeV) are not relevant because their contribution to the 
thermal axion potential is suppressed at the 
temperatures where the axion starts to 
oscillate, 
while the muon and light quark Yukawa operators predict an nEDM that exceeds the experimental bound.
    
For the electron Yukawa scenario (cf.~top panel of figure \ref{fig:summaryPlots}), we find that the associated PQ-breaking operator generates a scalar axion coupling 
to electrons which significantly constrains this scenario through long-range forces. These 
limits, in particular searches for fifth forces, 
imply that all of the observed DM can only be accounted for in the regime $m_a \gtrsim 2 \times 10^{-2}$ eV, which requires non-standard axion models to 
evade astrophysical limits on axion couplings to nucleons and electrons, also known as astrophobic axion models. 
This scenario is testable with a moderate improvement in fifth forces searches on length scales on the order of 5 $\mu$m.

For the $GG$ scenario (cf.~bottom panel of figure \ref{fig:summaryPlots}), we find that the axion scalar 
interaction with nucleons is (barely) dominated by the indirect contribution from $\thetaVEV$, which places this scenario in the standard regime in which the nEDM is the most sensitive probe. Therefore, long-range forces play no major role in this case. 
This scenario motivates QCD axion DM in the entire parameter range $10^{-5}$ eV $\lesssim m_a \lesssim 10^{-1} $ eV, if an astrophobic model is considered, or $10^{-5}$ eV $\lesssim m_a \lesssim 10^{-2} $ eV for a conventional KSVZ or DFSZ model.
Interestingly, although an $\mathcal{O}(1)$ part of the parameter space for the angles $\thetaini$ and $\deltaPQV$ is compatible with nEDM constraints, 
one generically predicts a signal in nEDM just beyond the current observational bound. This scenario is therefore also testable with a moderate improvement in the nEDM sensitivity.

Both of these realizations of trapped misalignment motivate axion DM with $m_a$ in the range of both high-mass haloscopes such as BREAD \cite{BREAD:2021tpx}, WISPFI \cite{Batllori:2023pap} and LAMPOST \cite{Baryakhtar:2018doz}, as well as the helioscope IAXO \cite{IAXO:2019mpb}. The reach of these experiments is indicated in figure \ref{fig:summaryPlots}.

Finally, we comment on avenues in which this study may be extended. Firstly, while our results predict QCD axion DM solutions tantalizingly close to current observational bounds, for the gluon scenario we rely on QCD dynamics at temperatures $T\sim 1 $ GeV near the QCD confinement scale. 
Our results are only to be trusted at the $\mathcal{O}(1)$ level, and it would be 
worth to improve on this.
Furthermore, in this study, we left out the photon coupling $\left(\phi/\LambdaUV\right)^n FF$. It would be interesting to extend this study with a careful treatment of the (2-loop) thermal effects 
and the associated constraints on this scenario. 
Finally, our setup relies on terms that couple the axion to a single SM operator without introducing a $\phi^n / \LambdaUV^{n-4}$ operator. We have taken a bottom-up approach and have not attempted to find a UV completion for this scenario. Clearly, it would be relevant to identify UV structures that naturally realize such a term.

\section*{Acknowledgments}

We thank 
Andreas Ekstedt,  
Hector Gisbert, 
Fabrizio Nesti, 
and Luca Vecchi
for useful discussions. 
This work is funded by the European Union -- NextGeneration EU and by the University of Padua under the 2021 STARS Grants@Unipd programme (Acronym and title of the project: CPV-Axion -- Discovering the CP-violating axion) as well as
by the European Union -- Next Generation EU and
by the Italian Ministry of University and Research (MUR) 
via the PRIN 2022 project n.~2022K4B58X -- AxionOrigins.  

\appendix

\section{On the origin of the \texorpdfstring{$\phi^n$}{phi-n} operators}
\label{app:model-building}
If the operator $\phi^n \mathcal{O}_{\rm SM}$ arises from a $Z_n$ symmetry realized at some UV scale $\LambdaUV$ then we would naively expect to also have terms of the form 
\begin{gather}
\label{eq:VLambdap}
    \mathcal{L}\supset e^{-i\delta'}\left(\frac{\phi}{\LambdaUV}\right)^n \LambdaPrime^4+\text{h.c.}\, ,
\end{gather}
where $\LambdaPrime$ is some energy scale and $\delta'$ is the misalignment angle with respect to the QCD potential. Naively, we would expect $\LambdaPrime \sim \LambdaUV$. 
Because this term is temperature-independent, it eliminates the advantage of the temperature-dependent potentials considered in section \ref{sec:Tdeppotentials}.
To understand how this potential may be pathological to our setup, consider the total axion potential
\begin{gather}
    V_{\rm tot} = - 2\left(\frac{\frac{1}{\sqrt{2}}f_a}{\LambdaUV}\right)^n\LambdaPrime^4\cos(n\theta+\delta') -\LambdaPQV^4(T) \cos(n\theta + \deltaPQV) + \VQCD \, .
\end{gather}
We can assume that the $\LambdaPQV(T)$ term dominates misalignment, such that the phenomenology of section \ref{sec:Tdeppotentials} is realized. However, at zero temperature, the $\LambdaPrime$ term implies an axion VEV of magnitude
\begin{gather}
    \thetaVEV \approx - 2n\sin(\delta')\frac{  \LambdaPrime^4 }{\chiQCD}\left(\frac{\frac{1}{\sqrt{2}}f_a}{\LambdaUV}\right)^n \, .
\end{gather}
Assuming that the axion makes up a DM fraction, $\dmfraction$, then $\LambdaUV$ is fixed according to eq.~\eqref{eq:LambdaUV DM sol, general}.
If we then impose the nEDM bound of $\thetaVEV \leq 10^{-10}$, 
and assume for definiteness the $GG $ coupling eq.~\eqref{eq:PQVGG}, we arrive at the upper bound   
\begin{gather}
    \LambdaPrime \leq 0.4 \text{ GeV } \frac{\abs{\thetaTrap}^{0.86}}{r^{0.43}\sin^{1/4}(\delta')} \, .
\end{gather}
Clearly, if the $\LambdaPrime$ term is present, then we must have a sharp hierarchy $\LambdaPrime \ll \LambdaUV$. This presents a 
model-building 
challenge, since it is not clear how a $\left(\phi/\LambdaUV\right)^n\mathcal{O}_{\rm SM}$ term can be motivated without generating the corresponding constant-temperature term with $\LambdaPrime\sim \LambdaUV$. This issue also affects the scenario in Ref.~\cite{Zhang:2023gfu}. We here choose to take a bottom-up approach in which 
the operator in eq.~\eqref{eq:VLambdap} is neglected, 
although we do point out the non-trivial task of finding a UV origin for the $\phi^n$ operators.

\bibliography{lib}

\end{document}